\documentstyle[aps,twocolumn,prl,epsfig]{revtex}

\newcommand{\be}{\begin{equation}}
\newcommand{\ee}{\end{equation}}
\newcommand{\ba}{\begin{eqnarray}}
\newcommand{\ea}{\end{eqnarray}}
\newcommand{\bi}[1]{\bibitem{#1}}
\newcommand{\fr}[2]{\frac{#1}{#2}}

\newcommand{\gsim}{\lower.7ex\hbox{$\;\stackrel{\textstyle>}{\sim}\;$}}
\newcommand{\lsim}{\lower.7ex\hbox{$\;\stackrel{\textstyle<}{\sim}\;$}}
\begin{document}

\twocolumn[\hsize\textwidth\columnwidth\hsize\csname @twocolumnfalse\endcsname

\preprint{
 RAL-TR-1999-082 \\
 UMN-TH-1832/99, TPI-MINN-99/60\\ hep-ph/9912293}

\title{ Theta angle versus CP violation in the leptonic sector}
\author{A. Dedes$^1$ and  M. Pospelov$^2$}
\address{ $^1$Rutherford Appleton Laboratory,
Chilton, Didcot, Oxon OX11 0QX, UK}
\address{$^2$Theoretical Physics Institute, School of Physics and Astronomy, 
University of Minnesota, Minneapolis, MN 55455, USA }

\maketitle
\begin{abstract}
Assuming that the axion mechanism of solving the strong CP 
problem does not exist and the vanishing of $\theta$ at tree
level is achieved by some model--building means, 
we study the naturalness of having large 
CP--violating sources in the leptonic sector. We consider the 
radiative mechanisms which transfer a possibly large CP-violating 
phase in the leptonic sector to the $\theta$ parameter. 
It is found that large $\theta$ cannot be induced in the models with
one Higgs doublet as at least three loops are required in this case.
In the models with two or more Higgs doublets the dominant source of 
$\theta$ is the phases in the scalar potential, induced by CP violation
in leptonic sector. Thus, in the MSSM framework the imaginary part 
of the trilinear soft-breaking parameter $A_l$ generates the 
corrections to the theta angle already at one loop. 
These corrections are large, excluding the possibility of
large phases, unless the universality in the slepton 
sector is strongly violated.

\end{abstract}

\vspace*{3mm}
]

\noindent

\section{Introduction and Motivation}

The strong CP problem whose existence was realized over twenty 
years ago \cite{theta} remains a complete mystery. 
The theta term of the QCD Lagrangian 
breaks P and CP invariance, and thus induces a variety of P,T-odd
observable effects, among which the electric dipole moments (EDMs) of
the neutron and heavy atoms play a prominent role \cite{KL}. 
The conflict between strong limits 
on $\theta$ resulting from  experimental 
searches of EDMs and natural expectations of $\theta\sim 1$ presents
a severe fine-tuning problem, usually referred to as the strong CP problem.
Using the experimental limits on the EDM of the neutron \cite{EDMn}
together with the result of a recent QCD sum rule calculation of 
$d_n(\theta)$ \cite{PR} one can place a very stringent limit on 
the theta term,
\begin{equation}
\theta< 6\cdot 10^{-10}.
\label{bound}
\end{equation}
A common and universal solution to the strong CP problem may come through
a dynamical relaxation mechanism \cite{PQ} which requires the existence 
of a light pseudoscalar (axion~\cite{WW})
 in the particle spectrum. Negative results from
experimental searches~\cite{PDB} of an axion together with very restrictive 
astrophysical~\cite{AB} and cosmological bounds~\cite{CB}
 on its coupling constant 
stimulate searches for alternative solutions. 

Another  possibility
is a model-building construction 
where $\theta$ can be naturally chosen to be zero at some high-energy scale 
due to exact parity or CP symmetry \cite{P,CP}. In this case however,
$\theta$ is not protected against radiative corrections at lower scales
where parity and/or CP symmetry are spontaneously broken. Thus the theta
term is extremely sensitive to the presence of additional, other than 
Kobayashi Maskawa, CP-violating sources in the hadronic sector. 
This sensitivity is unique: 
$\theta$ can receive contributions from the CP-violating phases
in the ``heavy'' sector of the theory without power-like suppression,
in contrast with other CP-violating operators. Thus, in the SUSY
variants of these models 
large soft-breaking phases in the squark and gluino sectors are 
excluded, as they
penetrate into the low-energy effective expression for $\theta$ 
already at one loop level. Therefore, a necessary consequence of 
these constructions seems to be a  strong restriction on CP violation, i.e.
no CP violation other than KM phase. Is this also true for CP violation 
which resides solely in the leptonic sector? In other words,
how  susceptible is $\theta$ to the CP-violation in the leptonic sector?

If the axion mechanism does not exist, the theta term 
is expected to be a dominant source of CP violation at 
low energy as it is the CP-odd operator of lowest dimension.
What would be a signal of the ``$\theta$-dominance'' among CP-violating 
observables? Both neutron and mercury EDMs produce similar bounds on 
$\theta$ and one should naturally expect that
\begin{eqnarray}
\label{dominance}
d_n \simeq 10^{-26}~ e\cdot cm ~\frac{\theta}{10^{-10}} 
\;\;\;\mbox{\cite{PR}}\nonumber\\
d_{Hg}\simeq 10^{-28}~ e\cdot cm ~\frac{\theta}{10^{-10}}
\;\;\;\mbox{\cite{PR}}
\label{EDMth}\\ 
d_{Tl} \sim 2\cdot10^{-29}~ e\cdot cm ~\frac{\theta}{10^{-10}} 
\;\;\;\mbox{\cite{nobody}}  \nonumber
\end{eqnarray}
Comparing the predictions of $\theta$-dominated EDMs with current 
experimental limits, \cite{EDMn,EDMHg,EDMe}, one can easily see that for
$\theta=10^{-10}$, $d_n$ and $d_{Hg}$ are within a factor of 2-3 from the
current experimental figures, whereas $d_{Tl}$ is  
smaller by five orders of magnitude than its present limit.
In other words, $\theta=10^{-10}$ will produce thallium EDM at the level 
equivalent to $d_{Tl}$, induced by the electron EDM 
$d_e\sim 4 \cdot 10^{-32}e\cdot cm$. Thus, it appears that the signal of 
$\theta$-dominance could be easily distinguished from the case of MSSM with 
large CP SUSY phases and axion-type solution to the strong CP problem. In 
the latter case $d_{Tl}$ is expected to be much more important than in
(\ref{EDMth}) and competitive with $d_n$ and $d_{Hg}$. 

However, if CP violation is initially concentrated in the leptonic sector,
the  ``$\theta$-signal'' (\ref{EDMth}) could be different. 
In this case the EDM of the electron and 
$d_{Tl}$ could be enhanced relative to
(\ref{EDMth}) and, at the same time, the $\theta$ term, induced by a
lepton CP-phase via radiative corrections would still dominate 
$d_n$ and $d_{Hg}$. 

The purpose of this note is to study the mechanisms of transferring 
CP violation from the leptonic sector to the theta term in the context 
of different models without an axion. 
Assuming no fine tuning which would compensate
an induced value of $\theta$, we find a ``maximal'' amount of CP-violation in 
the lepton sector, which can be consistent with the bound (\ref{bound}).
At the same time, we study possible enhancement of $d_e$ and $d_{Tl}$ 
due to the same sources of CP violation and the departure from the 
$\theta$-dominance signal, eq. (\ref{EDMth}).
 
\section{non-SUSY models}

We begin some remarks about the way how the low energy
value of $\theta$ should be calculated in a generic theory with CP-violation.
Besides the initial value of $\theta_{QCD}$, the relevant low energy parameter
$\bar\theta$ receives tree level contributions from the phases of the 
quark masses and other $SU(3)_c$-charged fermions. 
\be
\bar\theta=\theta_{QCD}+arg\,{\rm det}(M_u M_d)+...
\ee
It is often assumed in the literature, that the
radiative corrections to $\bar\theta$ are simply 
contained in the imaginary parts
of the quark and gluino masses. This is certainly true at the tree level, but
at the loop level the structure of radiative corrections is more complicated.
To give a simplest example, one can consider an effective Lagrangian
for gluons and quark field $q$ which arises after integrating out 
some unknown CP-violating physics at the scale $\Lambda$:
\begin{eqnarray}
{\cal L}_{eff}&=& \theta(\Lambda)\fr{g^2_3}{16\pi^2} G^a_{\mu\nu}
\tilde G^{a \mu\nu}+
\bar q (i \partial_\mu\gamma^\mu -m -im'\gamma_5)q - \nonumber
\\
&&\fr{im''}{2\Lambda^2}
\bar q  G^a_{\mu\nu}t^a\sigma^{\mu\nu}\gamma_5 q +...
\label{effective}
\end{eqnarray}
Here $\theta(\Lambda)$ denotes the theta term, coming from the scale 
$\Lambda$.
Let us take for simplicity $m\gg \Lambda_{QCD}$ and $m'\ll m$. 
Then the field $q$ can be also 
integrated out and the theta parameter below the scale $m$ reads as
\be
\bar\theta = \theta(\Lambda) + \fr{m'}{m} + 
\fr{m m''}{\Lambda^2}\log(\Lambda^2/m^2).
\ee
The second term in this expression is the ``usual'' correction 
due to the phase of the mass term, whereas the third term is 
generated by the ``chromoelectric dipole'' in (\ref{effective}). 
It is usually smaller than the second term due to $\Lambda\gg m$, 
although not necessarily negligible. For example, the scale 
of new physics $\Lambda$ could be comparable to the 
mass of heaviest fermions (top quark) so that the 
ratio $m m''/\Lambda^2$ is not small, or $m'$ can be simply 
zero from additional symmetry arguments and then the third term
dominates the expression for $\bar\theta$.
The latter is exactly the case in the minimal SM, 
where $\bar\theta$ recieves corrections from
``dipole'' contributions as it was first shown by Khriplovich \cite{Kh}. 
Technically, the corrections to $\bar\theta$ can be easily calculated 
within the external field formalism which will automatically account for all
contributions.

In what follows we determine possible mechanisms of transmitting
CP violation from the leptonic sector into the theta term in various possible
models \cite{BS}. As representative examples we take the 
Standard Model extended by right handed neutrino fields,
dilepton Zee model \cite{Zee}, multi-Higgs models
and the Minimal Supersymmetric Standard Model (MSSM) in particular.

It turns out that the main criterion which governs the efficiency of 
transmitting the CP violation from the leptonic sector into the theta term is
the number of weak doublets which give masses to the quark fields.
The contribution of the quark masses into $\bar\theta$ can be separated 
into the contributions of Yukawa couplings and Higgs  vevs:
\begin{eqnarray}
\label{decomp}
arg\,{\rm det}(M_u M_d) =
arg\,{\rm det}(Y_u)+ arg\,{\rm det}Y_d +\\ \nonumber 
3(arg \,v_u+arg \,v_d).
\end{eqnarray}
We take the vanishing of this expression due to some symmetry arguments
(for example, hermiticity of $Y_i$, reality of $v_i$'s) as the starting point
for our analysis. 

In the SM and in other models where $v_u\equiv v_d^*$, the contribution from 
the second line in (\ref{decomp}) is identically zero,
irrespective of the presence of CP-violation. Therefore the 
only way to insert CP-violation into the theta term is to ``complexify''  
quark Yukawa couplings and/or create quark chromoelectric dipoles.

Nontrivial corrections to quark Yukawa couplings sensitive to 
a CP phase in the leptonic sector
must be induced via Yukawa and $SU(2)\times U(1)$ gauge interactions. 
Furthermore, it is clear that in the presence 
of only one Higgs doublet  Yukawa interactions alone are not 
sufficient to achieve this. In any possible graph, involving a quark line 
and leptons in the loop, it is convenient to separate the loop part 
where actual CP violation takes place. Let us suppose now that
the particles circulating in the loop are 
heavy (Majorana neutrinos, for example)
and the lines, connecting the leptonic loop to a quark line are ``soft''.
Then it is possible to classify the effects of CP-violation in 
the leptonic sector in terms of effective CP-odd operators with dimension
6 and larger: $ H^\dagger H (B^{\mu\nu}\tilde B_{\mu\nu});\;
H^\dagger H (W^{a\mu\nu}\tilde W^a_{\mu\nu});\;
\tilde W^a_{\mu\nu} W^{b\nu\alpha} W^{c\mu}_\alpha\epsilon_{abc}$, etc.
One needs at least two loops to attach these operators to a quark line
with no external $SU(2)$ or $U(1)$ fields allowed. Together with at least
one (leptonic) loop needed to generate these operators, 
 three loops is the 
{\em minimal} order in which CP violation from the 
leptonic sector  penetrates into $\bar\theta$! 

In practice, the loop level is often  higher. In the SM  
 with heavy Majorana neutrinos, singlets of the SM gauge group,
one should have a minimum of four flavour-changing 
vertices on the lepton line. In the weak basis, which is more 
convenient because the momenta flowing in the loop are large, of the 
order of the heavy Majorana masses, these can only come from interactions
with the Higgs doublet. This adds another loop and
indicates that the effect may first appear at 
the four-loop order and a typical diagram is shown in Fig. 1.

\begin{figure}
\centerline{\psfig{figure=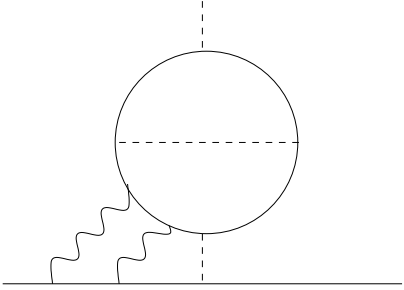,angle=0,height=1.5in}}

\vspace{0.5cm}

\caption{ A typical diagram, which gives a phase to the
quark Yukawa coupling. The circle is  ordinary leptons and Majorana
neutrinos, wavy lines are gauge bosons of the electroweak group and
dashed lines  the Higgs field. Similar diagrams work for the Zee model with
the dashed line inside the circle being a dilepton.}
\label{fig0}
\end{figure}

This diagram will be further suppressed by at least the 
square of the charged lepton mass as CP violation disappears if all 
charged leptons are massless. A more detailed calculation may reveal 
further suppression factors. For our purposes it is sufficient to 
acknowledge that the suppression factor is at least
\be
\bar\theta < \left({\alpha\over 4\pi}\right)^2 
 \left({1\over 16\pi^2}\right)^2 {m_\tau^2 \over M^2} J_{CP}^L,
\ee
where $M$ is the relevant high energy scale, at least as
heavy as $M_W$. No matter how large the CP violating combination
of mixing angles $ J_{CP}^L$ in the leptonic sector is, 
the result for $\bar\theta$
is well within the experimental bound. Therefore all CP violating phenomena
discussed in the literature such as CP violation in neutrino oscillations,
CP violation in the heavy Majorana neutrino decay, needed for leptogenesis
and others are entirely possible without causing problems for $\theta$. 

Precisely the same estimates (in this crude approach)
can be applied to the Zee 
model to produce similar conclusions, i.e. $\theta$ generated 
from the CP violation in the leptonic sector is small. However,
unlike the SM with Majorana neutrinos, where the possible 
electron EDM  is likely to be very small \cite{NgNg}, the Zee model
can have $d_e$ at a measurable level \cite{BS}. 

Another group of models has two or more Higgs doublets which give
masses to quarks via two or more {\em different} {\em v.e.v.}s.
In this case one should look for the effects which introduce phases
into the scalar potential. Thus the operators $H_u^i H_d^j \epsilon_{ij}$,
$(H_u^\dagger H_u) H_u^i H_d^j \epsilon_{ij}$, etc. may enter with 
complex coefficients which then can lead to $(arg \,v_u+arg \,v_d) \neq 0$.

In the non-supersymmetric framework a consideration of the radiative
corrections to the scalar potential are somewhat flawed. 
Indeed, the dimension 2
$H_u^i H_d^j \epsilon_{ij}$-proportional term enters in the Lagrangian
multiplied by some mass squared parameter. In the non-SUSY framework, at the 
radiative level, this parameter will be sensitive to  the square of the
cutoff $\Lambda^2$ which by itself requires fine-tuning to ensure 
the stability of the electroweak scale. Thus, we believe
that the question of induced phases in the soft-breaking sector
and $\theta$ cannot be solved without a specified framework which ensures 
the stability of the scalar potential. Thus we abandon non-SUSY
two Higgs doublet  models and 
take the case of MSSM where we study in 
detail the value of $\theta$ versus complex 
soft-breaking terms in the leptonic sector.

\section{MSSM with complex soft-breaking parameters in the lepton sector}

We  concentrate only on the leptonic sector
of the MSSM superpotential, i.e.,
\ba 
{\cal W} \supset \epsilon_{ab}({\bf Y}_e)_{ij}L_i^a H_1^b \bar{E}_j \;,
\label{super}
\ea
and the  soft breaking terms,
\ba
{\cal L}_{soft} \supset 
  &-& \epsilon_{ab}\biggl [({\bf A}_e{\bf Y}_e)_{ij}\widetilde{l_L}_i^a H_1^b 
\widetilde{e_R}^*_j + {\rm h.c} \biggr ] \nonumber \\ &-& \biggl [\mu B
\epsilon_{ab}H_1^a H_2^b + {\rm h.c} \biggr ]\;,
\label{soft}
\ea
where as usual $\tilde{e},\tilde{l}$ are the corresponding scalar
components of the chiral superfields 
$L,\bar{E}$ appeared in eq.(\ref{super}).
Let us assume universality of the soft trilinear couplings
at the GUT scale, ${\bf A}_e={\bf A}_\mu ={\bf A}_\tau$, and
one common phase $\phi_A$ associated with them.
We  consider the third generation of leptons, i.e., $A_{\tau}$
where the Yukawa couplings are large as
compared to those of the first and the second generation.
Then the renormalization group running of the imaginary 
part of the parameter $A_{\tau}$, denoted
as $\bar{A}_\tau$, induces an imaginary part of the
parameter $B$, denoted as $\bar{B}$, at a scale below the GUT scale and
their RGEs are given by~\cite{martin},
\ba
\frac{d \bar{A}_\tau}{dt} \ &=&\ 
\frac{8 |Y_\tau|^2}{16\pi^2} \bar{A}_\tau \;, \\
\frac{  d\bar{B}}{dt} \ &=& \ \frac{2 |Y_\tau|^2}{16\pi^2} \bar{A}_\tau
\;.
\label{rge}
\ea
All the other parameters of the SUSY or the soft SUSY breaking sector 
remain real.
The  tau lepton Yukawa coupling has a weak
running (especially for small  values of $\tan\beta$).
 Thus the system of differential 
equations of (\ref{rge}) can be solved trivially and gives,
\ba
\bar{A}_{\tau}(Q) \ &=& \bar{A}_\tau (M_G) \biggl (\frac{Q}{M_G} 
\biggr )^{\frac{|Y_\tau|^2}{2\pi^2}} \;, \label{5} \\
\bar{B}(Q) \ &=& \  - 
\frac{\bar{A}_\tau(M_G)}{4} 
\biggl [ 1- \biggl ( \frac{Q}{M_G} \biggr )^{\frac{|Y_\tau|^2}{2\pi^2}}
\biggr ] \nonumber \\ \;. \label{sol}
\ea
So even if all the parameters at the GUT scale are real
apart from the leptonic trilinear coupling  
 i.e., $\bar{A}_\tau$, then this parameter affects the running 
of the $\bar{B}$ parameter and generates a non-zero $\bar{B}$.
We can easily see from (\ref{sol}) that the running of
the phase of the parameter $B$, i.e., $\phi_B$ at a scale Q 
is given by,
\ba
\sin\phi_B(Q)  =  -\frac{1}{4} \frac{|A_\tau(M_G)|}{|B(Q)|}\sin\phi_A(M_G)
 \biggl [ 1- \biggl ( \frac{Q}{M_G} \biggr )^{\frac{|Y_\tau|^2}{2\pi^2}}
\biggr  ]   \nonumber \\  
\;\;\;\;\;\;\;\;\;\;\;\;\;.
\label{phaseB} 
\ea

Let us now see what happens at the EW scale i.e., $Q=M_Z$.
It is reasonable to take $|A_\tau(M_G)| \simeq |B(M_Z)|$ in the 
MSSM with Radiative Electroweak breaking. This assumption of course
depends on the choice of the other MSSM  parameters, $M_0$, $M_{1/2}$,
$A_0$ and $\tan\beta$. We
 display the numerical solutions below.
For $|Y_\tau|^2/4\pi \simeq 4\times 10^{-5}$ and $\tan\beta=2$
with $M_{GUT}=3\times 10^{16}$ GeV we get, from eq.(\ref{phaseB})
\ba
\sin\phi_B(M_Z) \simeq -2\times 10^{-4} \sin\phi_A \;.
\label{eq10}
\ea

Now from the minimization conditions of the scalar 
Higgs potential we have,
\ba
v_1 v_2 \ =\ \frac{\mu^* B^* |v|^2}{m_1^2 + m_2^2} \;,
\ea
where $m_{1,2}^2 = m_{H_{1,2}}^2 +\mu^2$ and $|v|^2=|v_1|^2 + |v_2|^2$.
Note also that the parameter $\mu$ remains real (if originally is real)
at every scale because its renormalization is multiplicative.

The $\theta$ angle is generated if $B$ is complex and
given by eq. (\ref{decomp}).
Putting the experimental bound (\ref{bound}) of $\theta$
parameter  into eq.(\ref{eq10}) we get 
\ba
\phi_A(M_G) < 10^{-6} \;,
\label{constraint}
\ea
an unnatural small number at the GUT scale. 
We conclude that the phase in the leptonic sector 
produces large additive renormalization of $\theta$-QCD parameter
which constitutes a fine tuning problem 
unless this phase is tiny, of the order of $10^{-6}$ or smaller. 

Three remarks are in order: i) Even if we assume an appropriate
phase for the parameter $\mu$ at a scale $Q$ which cancels
the contribution of the $\phi_B$ i.e., $\phi_\mu(Q)=-\phi_B(Q)$
then eventually this pattern will be  destroyed by the running of $\phi_B$ 
of eq.(\ref{sol}) since $\phi_\mu$ does not run, ii)
the  constraint (\ref{constraint}) on $\phi_A(M_G)$ is relaxed if
we consider non-universality of the soft SUSY breaking trilinear
couplings at the GUT scale in 
the case of the electron and muon Yukawa couplings, iii) if
we start with the (trivial) case  $A=0$~GeV at the GUT scale then there is
no contribution to the $\theta$-term and no CP-violation in the
leptonic sector. 

\begin{figure}
\centerline{\psfig{figure=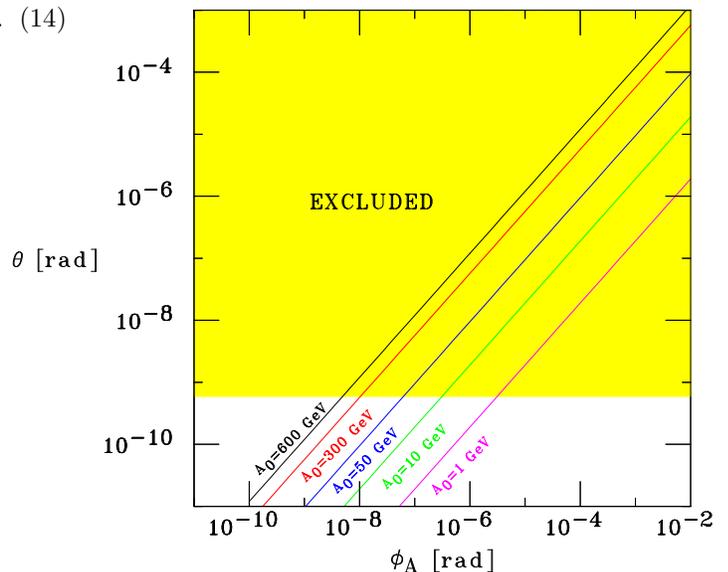,angle=90,height=3in}}
\caption{The extracted value of the $\theta$-term as a
function of the common phase $\phi_A$ at the GUT scale  of the 
lepton trilinear soft SUSY breaking couplings. The other SUSY
breaking parameters have been fixed  $M_0=M_{1/2}=200$~GeV and
$\tan\beta=10$. The shaded region is excluded by the experiment, 
 {\it see} (\ref{bound}). Results
on $\theta$ from 
different values of the modulo $|A_0|=1,10,50,300,600$~GeV are
also indicated. }
\label{fig1}
\end{figure}

We perform a numerical analysis of the RGEs by also taking
 into account low energy threshold effects~\cite{sakis}.
We present our results in Fig.~\ref{fig1}. We see
that  $\phi_A \lsim 10^{-6}$ unless $A_0$ is
exactly zero. Small departures from  zero 
(see the line with $|A_0|=1$~GeV in Fig.(\ref{fig1}) for
instance) put a strong bound on the phase $\phi_A$.
As $|A_0|$  increases the bound becomes stronger; as strong as
$\phi_A \lsim 10^{-8}$ for $|A_0| \gsim 300$~GeV. 
This happens because  $|A_0|=|A_\tau(M_G)|$ gets
much larger than  $|B(M_Z)|$ which further enhances the value of theta,
as  seen from eq.(\ref{phaseB}).

Therefore we  face two possible choices if we still 
want to keep large CP-violation in the leptonic sector: i)
relax the universality pattern of the
phases at the GUT scale (however, even in that
case the phases of the $\tau$ trilinear soft
breaking coupling must be unnaturally small as
we prove above) ii) introduce PQ symmetry and the axion 
solution to the strong CP problem.

\section{Conclusions}
We have studied the question of naturalness for CP violation in the 
leptonic sector to be  large without inducing large 
corrections to $\bar\theta$. This is an important question in the
context of non-axionic solutions to the strong CP problem.
We find that the main criterion dividing  models into two  classes
is the number of Higgs doublets giving masses to quarks. In the case of 
one doublet the contribution of CP phases from the leptonic sector to
$\bar\theta$ is always small, being suppressed by at least a three-loop
factor so that a ``maximal'' CP violation in the leptonic sector is
allowed. In some of these models (dilepton model, for example), 
$d_e$ can be quite large, enhancing $d_{Tl}$ with respect to 
``$\theta$-dominance'' signal, eq. (\ref{dominance}), usually expected when 
the axion mechanism is absent.

In the models with several doublets there is an efficient way of 
transmitting CP violation from the leptonic sector into $\theta$ via
complex parameters in the scalar potential. In MSSM without an axion, 
a large phase of the leptonic $A_l$-parameter is excluded on the ground 
of naturalness, unless the lepton universality is broken in a peculiar way
that only $A_e$ (or $A_\mu$) has the phase. 

\vspace{2.0mm}
\noindent
{\small{\sl We thank R. R. Roberts for valuable discussions.
 AD acknowledges financial support from 
 the Marie Curie Research Training Grant
ERB-FMBI-CT98-3438.
This work was supported in part by the Department of Energy
under Grant No DE-FG-02-94-ER-40823. }}


\begin{thebibliography}{99}

\bi{theta} C. Callan, R. Dashen and D. Gross, Phys. Lett. 
{\bf B63} (1976) 334, 
R. Jackiw and C. Rebbi, Phys. Rev. Lett. {\bf 37} (1976) 172. 

\bi{KL} I.B. Khriplovich and S.K. Lamoreaux, {\it ''CP Violation 
Without Strangeness''}, Springer, 1997. 

\bi{EDMn} K.F. Smith {\em et al.}, Phys. Lett. {\bf B234} 191 (1990);
I.S. Altarev {\em et al.}, Phys. Lett. {\bf B276} 242 (1992);
 P.G. Harris {\em et al.},  Phys. Rev. Lett. {\bf 82} 904
(1999).

\bi{PR} M. Pospelov and A. Ritz, Phys. Rev. Lett. {\bf 83} (1999) 2526; 
 M. Pospelov and A. Ritz, hep-ph/9908508.

\bi{PQ} R.D. Peccei and H. Quinn, Phys. Rev. Lett. {\bf 38} (1977) 1440; 
Phys.\ Rev.\ {\bf D16} (1977) 1791;
J.~E.~Kim, Phys. Rev. Lett. {\bf 43} (1979) 103;
M.A.~Shifman, A.I.~Vainshtein and V.I.~Zakharov,
Nucl.\ Phys.\ {\bf B166}, 493 (1980);
A.R.  Zhitnitsky, Sov. J. Nucl. Phys. {\bf 31} (1980) 260; 
\\ M. Dine,  W. Fischler and M. 
Srednicki, Phys. Lett. {\bf B104} (1981) 199.

\bi{WW}S.~Weinberg, Phys.\ Rev. \ Lett. \ {\bf 40,} (1978) 223;
F.~Wilczek, Phys.\ Rev. \ Lett. \ {\bf 40,} (1978) 279.

\bi{PDB} C.~Hagmann {\it et.al}, in Eur. \ J. \ {\bf C3,} (1998) 1 .

\bi{AB} For reviews, see M.~S.~Turner, Phys. \ Rept. \ {\bf 197,}
(1990) 67 ; G. G. Raffelt, Phys. \ Rept. \ {\bf 198,} (1990) 1 .

\bi{CB}J.~Preskill, M.B.~Wise and F.~Wilczek,
Phys.\ Lett.\ {\bf B120} (1983) 127;
L.F.~Abbott and P.~Sikivie,
Phys.\ Lett.\ {\bf B120} (1983) 133;
M.~Dine and W.~Fischler,
Phys.\ Lett.\ {\bf B120} (1983) 137. 
For reviews, see J.~E.~Kim, Phys. \ Rept. \ {\bf 150,} (1987) 1;
M.~S.~Turner, Phys. \ Rept. \ {\bf 197,} (1990) 68.



\bi{P} M.A.B. Beg and H.S. Tsao, Phys. Rev. Lett. {\bf 41} (1978) 278;
R.N. Mohapatra and G. Senjanovic, Phys. Lett. {\bf B79} (1978) 278.

\bi{CP}A.~Nelson, Phys.\ Lett.\ {\bf B136} (1984) 387; 
S. Barr, Phys. Rev. Lett. {\bf 53} (1984) 329; 
P.H. Frampton, Phys. Rev. Lett. {\bf 68} (1992) 2129; 
H. Georgi and  S.L. Glashow, Phys. Lett. {\bf B451} (1999).

\bi{nobody} We are not aware of any work which calculates an EDM of
a paramagnetic atom, $d_{Tl}$ or $d_{Cs}$, induced by the theta term.
One way to estimate the effect (M. Pospelov, 1992, unpublished) is to 
determine a size of the effective T-odd nucleon-electron interaction
$\bar NN \bar e\gamma_5e$, induced by $\theta$ via a 
$\pi_0$-exchange. In this case CP violation resides in the 
$\pi NN$ vertex, 
R. Crewther {\em et al.}, Phys. Lett. {\bf B88} (1979) 123, and 
$\bar e \gamma_5  e \pi_0$ interaction can be extracted from the $\pi_
0\rightarrow e^+e^-$ branching ratio. The final estimate is given in 
the third line of eq. (\ref{EDMth}).

\bibitem{EDMHg}  J.P. Jacobs {\em et al.}, Phys. Rev. Lett. {\bf 71} (1993) 
3782. 

\bibitem{EDMe}  E.D. Commins {\em et al.}, Phys. Rev. {\bf A50} (1994) 2960.
 
\bibitem {Kh} I.B. Khriplovich, Yad. Fiz. {\bf 44} (1986)
1019 (Sov. J. Nucl. Phys. {\bf 44} (1986) 659). The value of $\theta$, induced
by KM phase $\delta$ at the electroweak threshold is much bigger than the
renormalization group mixing of $\theta$ and $\delta$ considered in 
J. Ellis and M. Gaillard, Nucl. Phys. {\bf B150} (1979) 141.

\bibitem{BS} W. Bernreuther and M. Suzuki, Rev. Mod. Phys. {\bf 63}, 
313 (1991).

\bibitem{Zee} A. Zee, Phys. Rev. Lett. {\bf 55}, 2382 (1985).

\bibitem{NgNg} D. Ng and J. Ng, Mod. Phys. Lett. {\bf A11}, 211 1996. The
two-loop diagrams considered in this paper have additional suppression 
factors driving $d_e$ below the level of $10^{-32}e\cdot cm$.   

\bibitem{martin}
S.~P.~Martin and M.~T.~Vaughn, Phys. Rev. D {\bf 50}, 2282 (1994);
Y.~Yamada,
Phys.\ Rev.\ {\bf D50} (1994) 3537
hep-ph/9401241;
I.~Jack and D.R.~Jones,
Phys.\ Lett.\ {\bf B333} (1994) 372
hep-ph/9405233.

\bibitem{sakis}
A.~Dedes, A.B.~Lahanas and K.~Tamvakis,
Phys.\ Rev.\ {\bf D53} (1996) 3793
hep-ph/9504239.


\end{thebibliography}
\end{document}